\documentclass[iop,tighten,twocolumn,apj]{emulateapj}
\usepackage{multirow}
\usepackage{amssymb}
\usepackage{graphicx}
\usepackage{amsmath}
\usepackage{cprotect} % for verbatim in caption
\usepackage{paralist}
\usepackage{bm}% bold math.
\usepackage[breaklinks,colorlinks,citecolor=blue]{hyperref}

\begin{document}

\title{Ultra-diffuse and Ultra-compact Galaxies in the Frontier Fields Cluster
Abell 2744}

\author{Steven~Janssens\altaffilmark{1}, Roberto~Abraham\altaffilmark{1},
Jean~Brodie\altaffilmark{2}, Duncan~Forbes\altaffilmark{3},
Aaron~J.~Romanowsky\altaffilmark{2,4}, and Pieter~van~Dokkum\altaffilmark{5}}

\altaffiltext{1}{Department of Astronomy and Astrophysics, University of
Toronto, 50 St. George Street, Toronto, ON, Canada M5S 3H4}
\altaffiltext{2}{University of California Observatories, 1156 High Street,
Santa Cruz, CA 95064, USA}
\altaffiltext{3}{Centre for Astrophysics and Supercomputing, Swinburne
University, Hawthorn VIC 3122, Australia}
\altaffiltext{4}{Department of Physics and Astronomy, San Jos\'e State
University, One Washington Square, San Jose, CA 95192, USA}
\altaffiltext{5}{Department of Astronomy, Yale University, 260 Whitney Avenue,
New Haven, CT 06511, USA}

\email{janssens@astro.utoronto.ca}

\shorttitle{UDGs and UCDs in A2744}
\shortauthors{Janssens et al.}

\begin{abstract}
We report the discovery of a large population of Ultra-diffuse Galaxies (UDGs)
in the massive galaxy cluster Abell 2744 ($z=0.308$) as observed by the Hubble
Frontier Fields program. Since this cluster is $\sim5$ times more massive than
Coma, our observations allow us to extend 0.7 dex beyond the high-mass end of
the relationship between UDG abundance and cluster mass reported by
\cite{vdb2016}. Using the same selection criteria as \cite{vdb2016}, A2744
hosts an estimated $2133 \pm 613$ UDGs, ten times the number in Coma. As noted
by \cite{lee2016}, A2744 contains numerous unresolved compact objects, which
those authors identified predominantly as globular clusters.  However, these
objects  have luminosities that are more consistent with ultra-compact dwarf
(UCD) galaxies.  The abundances of both UCDs and UDGs scale with cluster mass
as a power law with a similar exponent, although UDGs and UCDs have very
different radial distributions within the cluster. The radial surface density
distribution of UCDs rises sharply toward the cluster centre, while the
surface density distribution of the UDG population is essentially flat.
Together, these observations hint at a picture where some UCDs in A2744 may
have once been associated with infalling UDGs. As UDGs fall in and dissolve,
they leave behind a residue of unbound ultra-compact dwarfs.
\end{abstract}

\section{Introduction}

It is now known that the Universe is not nearly as deficient in massive low
surface brightness galaxies as was once thought, and that such `ultra-diffuse
galaxies' (UDGs) can be found in large numbers in rich clusters of galaxies
\citep{vandokkum2015a,vandokkum2015b,koda2015,vdb2016}.  The largest
UDGs have sizes similar to the Milky Way (half-light radii
around 3 kpc) but only 1/100 to 1/1000 as many stars.
These systems were originally discovered using the Dragonfly Telephoto Array
\citep{abraham2014}, which is highly optimized for the detection of low
surface brightness structures, but the detection of most UDGs is within the
capability of conventional telescopes.

The discovery of UDGs has generated tremendous interest in the community, from
observers who are rapidly enlarging the UDG samples
\citep[e.g.][]{koda2015,mihos2016,vdb2016}, from simulators who must now try
to understand the origin and evolution of these galaxies
\cite[e.g.][]{yozin2015,amorisco2016a}, and even from alternative gravity
researchers who claim their existence challenges dark matter models
\citep{milgrom2015}. The existence of so many presumably `delicate' UDGs in
rich clusters (\cite{koda2015} put their number at $\sim 800$ in Coma) poses
the immediate question of why they are not being ripped apart by the tidal
field of their host clusters. They may be short-lived and be on their first
infall and about to be shredded, but this seems unlikely given their
predominantly red stellar populations and smooth morphologies.
However, two UDGs in Virgo show extended tidal debris and appear to be in the
process of being tidally stripped \citep{mihos2015,mihos2016}.
If they have survived for several orbits in a rich cluster, then simple
stability arguments suggest that they must have significantly higher masses
than implied by their stellar populations; in fact, in order to survive, their
dark matter fractions need to be $>98\%$ \citep{vandokkum2015a} within their
half-light radii, suggesting they are `failed' $L^\ast$ galaxies. At least two
objects in Coma, Dragonfly 17 \citep{peng2016} and Dragonfly 44
\citep{vandokkum2016} show strong evidence (high velocity dispersions or large
globular cluster populations, or both) for being resident within very massive
halos. So for at least two UDGs the `failed giant' picture appears to be
plausible. However, these may be extreme cases
\citep{amorisco2016b,roman2016b}, with more typical UDGs being better
described as `inflated dwarfs', whose anomalously large sizes are due to
extreme feedback-driven outflows \citep{dicintio2017}, unusually high spins
\citep{amorisco2016a}, or tidal disruption \citep{collins2014}.  At
present, very little is known about the characteristics of UDGs, and it is not
clear what fraction of them are `failed giants', `inflated dwarfs', or some
other phenomenon.

Another relatively newly discovered population of low-mass objects lies at the
opposite end of the selection function from UDGs. These `ultra-compact dwarfs'
(UCDs) have characteristics reminiscent of both the nuclei of low-mass
galaxies \citep{georgiev2014}, and massive globular clusters (GCs), and they
may well have a connection to both populations \citep[see,
e.g.,][]{mieske2002,mieske2012,brodie2011,norris2014,zhang2015}. UCDs seem to
occur mostly in dense environments (both near the centres of clusters and near
massive galaxies), suggesting that environmental factors (e.g.\ tidal
stripping) drives their formation \citep[e.g.,][]{bekki2003,pfeffer2013}.

With an eye towards better understanding the nature of both UDGs and UCDs, in
this paper we investigate the `extreme' galaxy populations in Abell 2744 using
data obtained with the {\em Hubble Space Telescope} (\textit{HST}) Frontier
Fields (FF) program. A2744, also known as the Pandora Cluster, is one of the
most massive \citep[virial mass $\sim 5 \times
10^{15}~\mathrm{M}_{\odot}$,][]{boschin2006,medezinski2016} and most disturbed
galaxy clusters known \citep{owers2011}.
Its intracluster light fraction is high at $19 \pm 3\%$
\citep{jimenezteja2016}, with a mass surface density of $\sim
10~\mathrm{M}_{\odot}~\mathrm{pc}^{-2}$ and a stellar population consistent
with the disruption of $L^\ast$ galaxies \citep{montes2014}. These properties
suggest A2744 is an ideal location to search for UDGs and UCDs at a look-back
time of $\sim$3.5 Gyr and we seek to learn whether its extreme characteristics
may have left an imprint in its population of UDG and UCD galaxies.

\begin{figure*}
	\includegraphics[width=\textwidth]{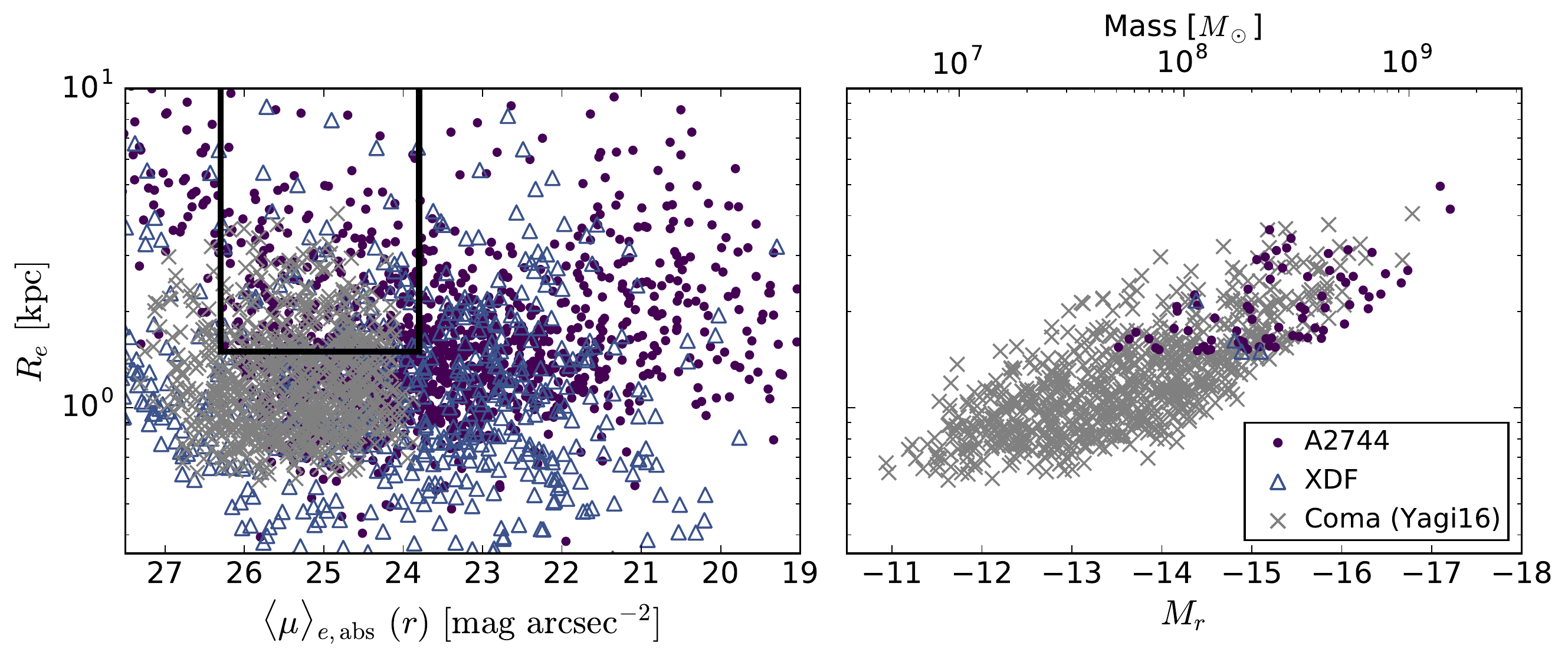}
	\caption{
	\textit{Left}: GALFIT circularized effective radii and the absolute
	mean surface brightness within $R_e$ of extended objects in A2744
	(cluster and parallel fields, purple dots) and the XDF (blue
	triangles), as well as Coma UDGs from \cite{yagi2016} (grey crosses).
	We select UDGs with $R_e \geq 1.5~\mathrm{kpc}$,
	$23.8 \leq \langle\mu\rangle_{e,\mathrm{abs}} \leq
	26.3~\mathrm{mag}~\mathrm{arcsec}^{-2}$ and S\'ersic index $n \leq 4$. 
	\textit{Right}: Sizes and absolute magnitudes, along with
	corresponding stellar masses, of visually-checked UDGs.
	\label{fig:galfit}
	}
\end{figure*}

In this paper, we adopt a $\Lambda$CDM cosmology with $\Omega_m = 0.3$,
$\Omega_{\Lambda} = 0.7$, $H_0 =
70~\mathrm{km}~\mathrm{s}^{-1}~\mathrm{Mpc}^{-1}$, and a redshift for A2744 of
$z = 0.308$, which corresponds to $m-M = 41.02$ and $1~\mathrm{arcsec} =
4.536~\mathrm{kpc}$.  All magnitudes are in the AB system. Galactic extinction
corrections from the \cite{schlafly2011} extinction maps were applied to all
magnitudes.\footnote{Using the online calculator at
\url{https://ned.ipac.caltech.edu/forms/calculator.html}.}

\section{Data}

The \textit{HST} FF program has produced the deepest images to date of galaxy clusters
and gravitational lensing for six clusters 
along with six parallel blank fields offset from each cluster
\citep{lotz2016}. Each cluster and parallel field were observed for 70 orbits
with the Advanced Camera for Surveys (ACS) in F435W, F606W and F814W, and 70
orbits with the Wide Field Camera 3 (WFC3) in F105W, F125W, F140W and F160W.
The filters with the deepest coverage are F814W, F105W and F160W. We use the
higher resolution 30 mas scale v1.0 images with the ``self-calibration"
applied to the ACS images and the time variable sky correction applied to the
WFC3 images (Koekemoer et al., in prep). The 30 mas images properly sample the
ACS point spread function (PSF).

We note that despite being offset $6\arcmin$ west of A2744's core, the
parallel field is well within A2744's virial radius \citep[$R_{200} = 9\arcmin
= 2.5~\mathrm{Mpc}$,][]{medezinski2016}.  To supply background corrections, we
relied on the \textit{HST} eXtreme Deep Field \citep[XDF,][]{illingworth2013}.
This is the deepest image of the sky to date in the optical/near-IR, and was
obtained by stacking data from 19 different \textit{HST} programs spanning 10
years covering the Hubble Ultra-Deep Field. The XDF has ACS coverage in F435W,
F606W, F775W, F814W and F850LP, and WFC3 coverage in F105W, F125W, F140W and
F160W.  Only 60 mas scale images are available for all filters.

\section{Methodology}

\subsection{Object Detection}

For the A2744 cluster and parallel field, we ran \textsc{SExtractor} 
\citep{bertin1996} in dual image mode on the 30 mas images using the F814W
image as the detection image for all bands.  To detect extended low
surface-brightness objects, \texttt{DETECT\_MINAREA} was
set to 20 pixels, and \texttt{DETECT\_THRESH} and \texttt{ANALYSIS\_THRESH}
were both set to 0.7 times the background RMS. Backgrounds were measured in
local annuli 24 pixels thick.

The XDF's F814W depth is relatively shallow, so instead we used F775W as the
detection band. The 60 mas pixels are 4 times the area so
\texttt{DETECT\_MINAREA} was set to 5 pixels.

\subsection{Point Spread Functions}

\textsc{PSFEx} \citep{bertin2011} was used to fit the PSF across the F814W
A2744 cluster and parallel field images. Stars were selected from a more
conservative \textsc{SExtractor} catalog with $\mathtt{DETECT\_MINAREA} = 5$
and $\mathtt{DETECT\_THRESH} = 1.0$ using the cuts $1.0 < \mathrm{FWHM} < 6.0$
pixels, $\mathrm{S/N} > 5$ and $e < 0.3$.  For the XDF, we again used the
F775W image with the same parameters as the A2744 FF, except we used
$\mathrm{FWHM} < 3.1$ pixels to select the PSF stars.
 
\subsection{Ultra-diffuse Galaxy Selection} \label{sec:udg_selection}

UDG candidates were selected based on their half-light radii and peak
surface-brightness.  Our approach is essentially that adopted by
\cite[][hereafter vdB16]{vdb2016}, with minor adaptations needed to account
for the fact that our observations are based on data obtained with
\textit{HST}. In brief, we followed a four-stage process: (1)
Low-surface brightness candidates were selected using
\textsc{SExtractor}. (2) Candidates were then filtered on the basis of
colour to isolate systems with rest-frame colours consistent with
quiescent galaxies at the redshift of A2744.  (3) Structural
parameters were obtained for the remaining candidates in order to
extract systems with sizes larger than 1.5 kpc, absolute $r$-band mean
surface brightnesses between $23.8 \leq \langle\mu\rangle_{e,\mathrm{abs}}
\leq 26.3~\mathrm{mag}~\mathrm{arcsec}^{-2}$ and S\'ersic index $n \leq
4$.  (4) Obvious image artifacts were discarded using visual
inspection. We now describe each of these four steps in some detail.

\begin{figure*}
	%A2744-2747
	\includegraphics[width=0.50\textwidth]{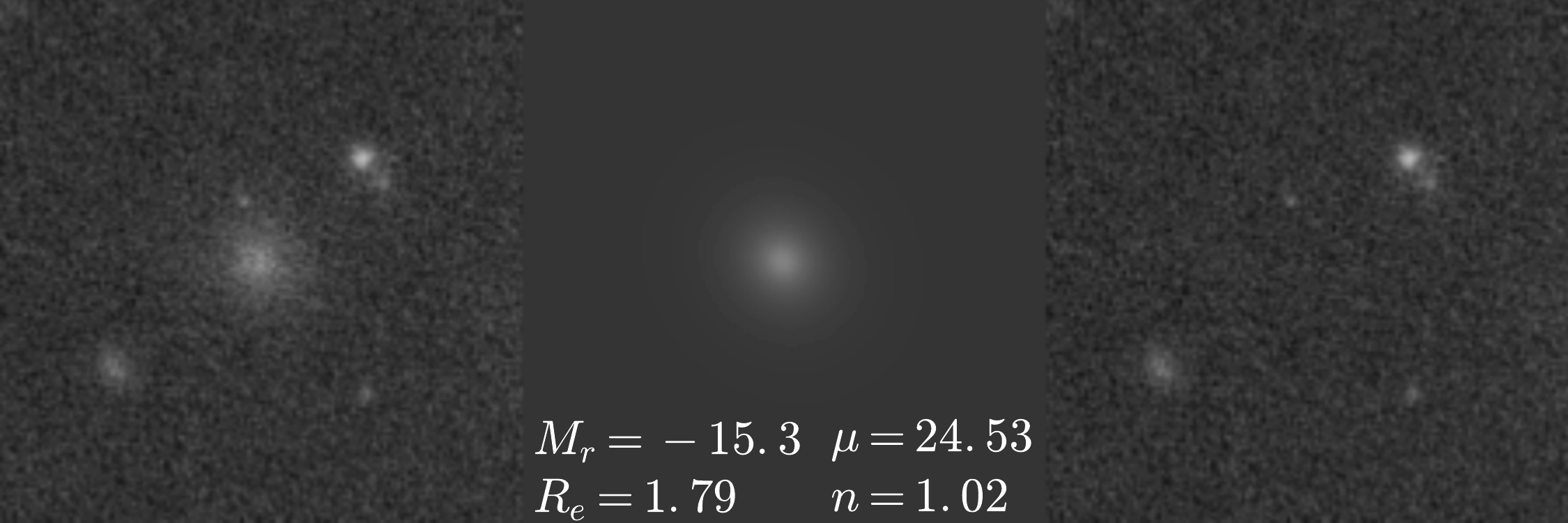}
	%A2744-2864
	\includegraphics[width=0.50\textwidth]{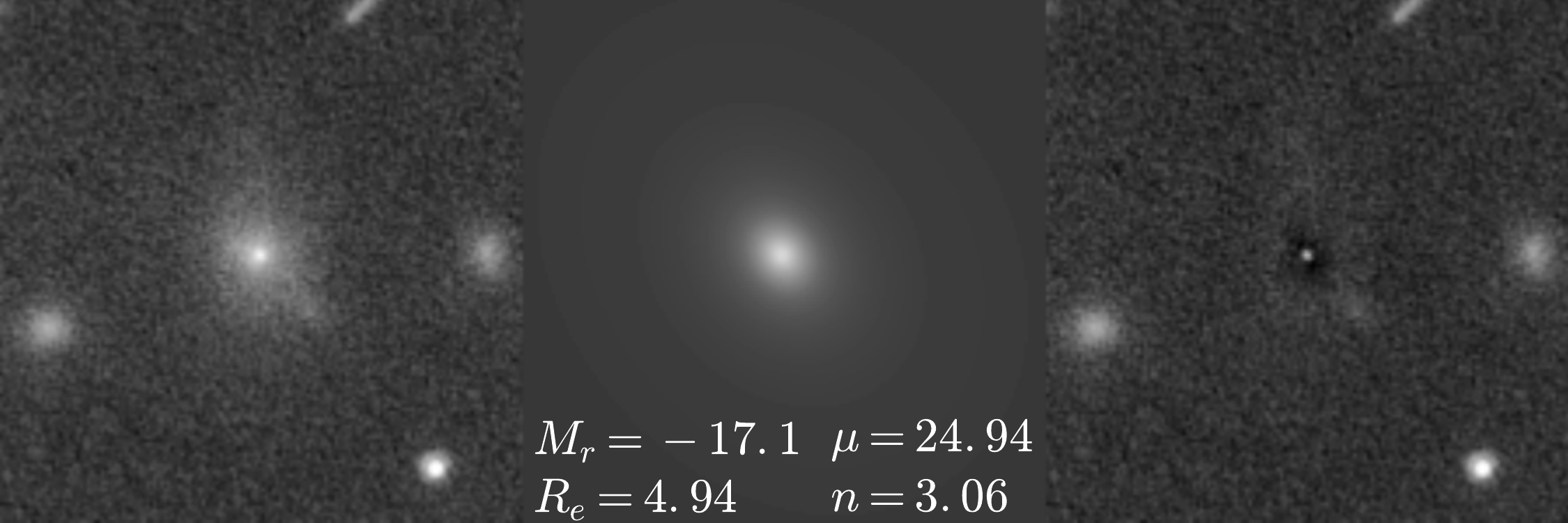}
	%A2744-4743
	\includegraphics[width=0.50\textwidth]{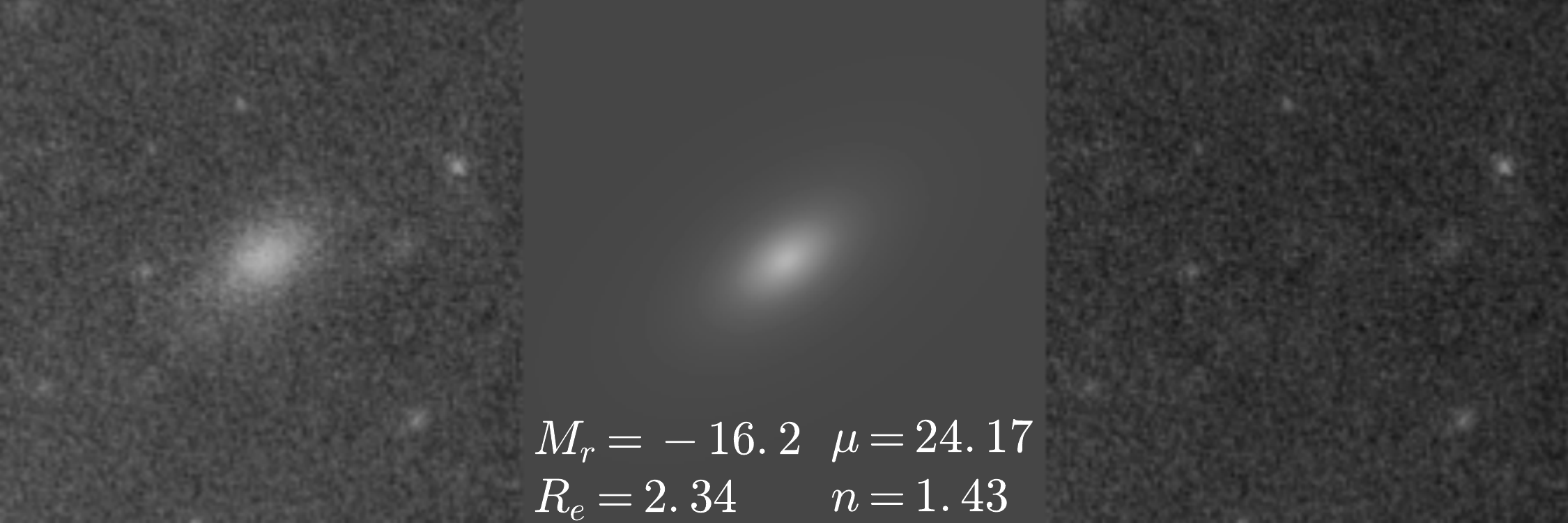}
	%A2744PAR-4653
	\includegraphics[width=0.50\textwidth]{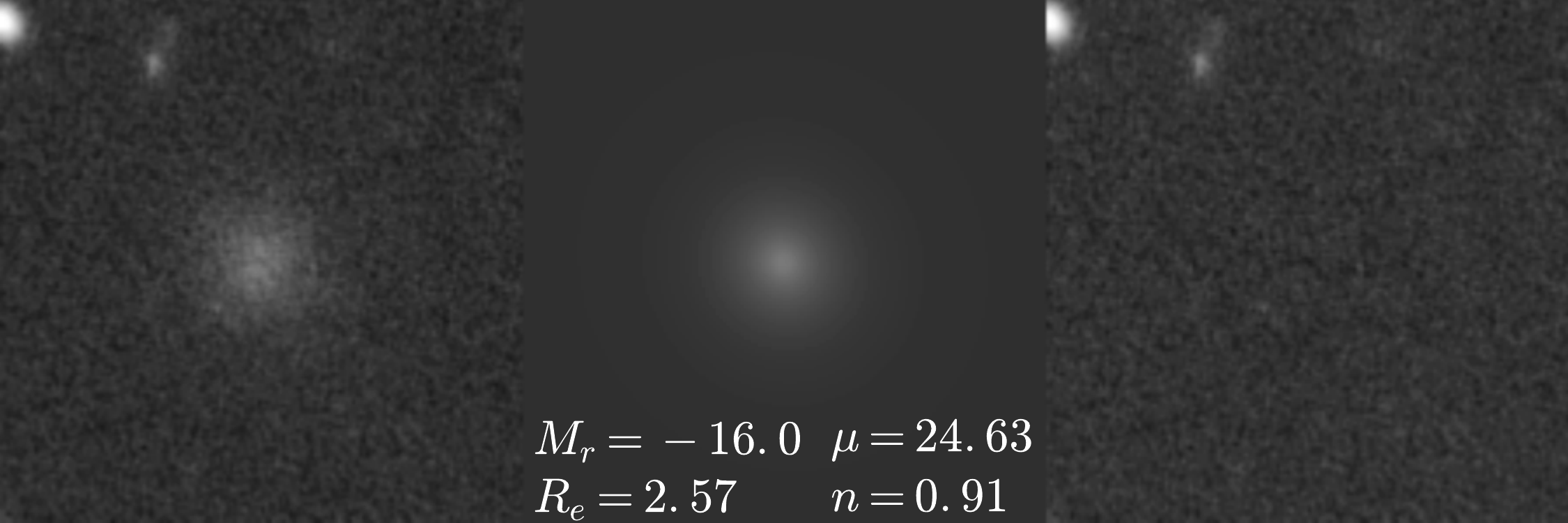}
	%A2744PAR-4807
	\includegraphics[width=0.50\textwidth]{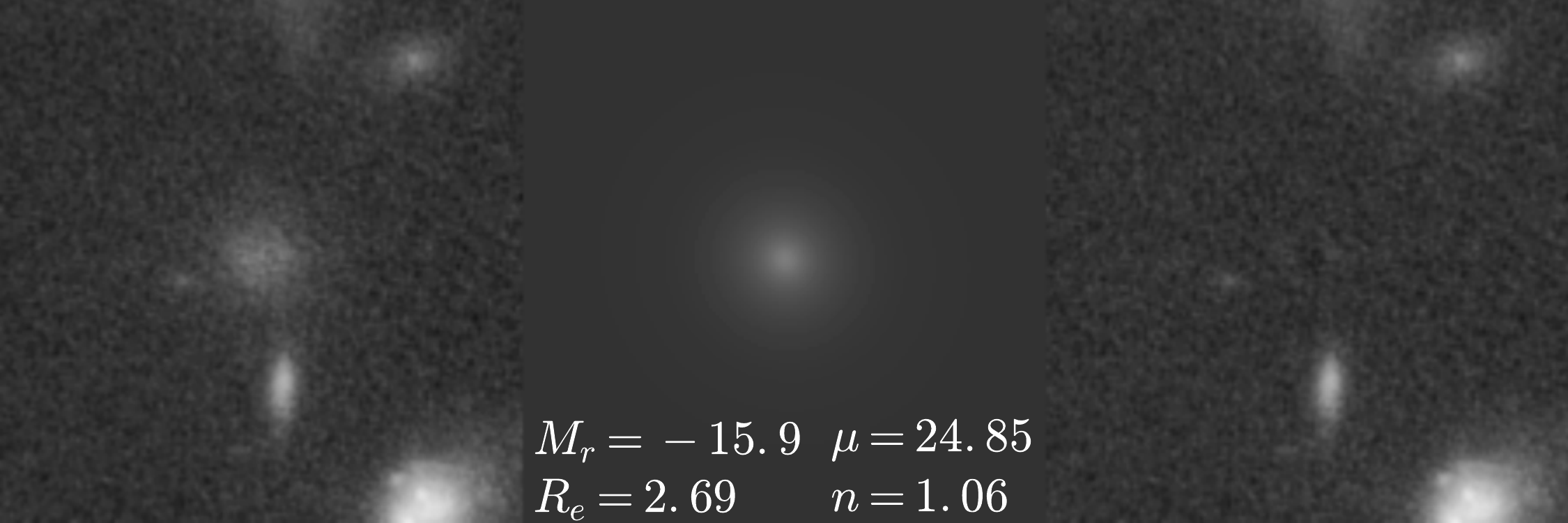}
	%A2744PAR-6341
	\includegraphics[width=0.50\textwidth]{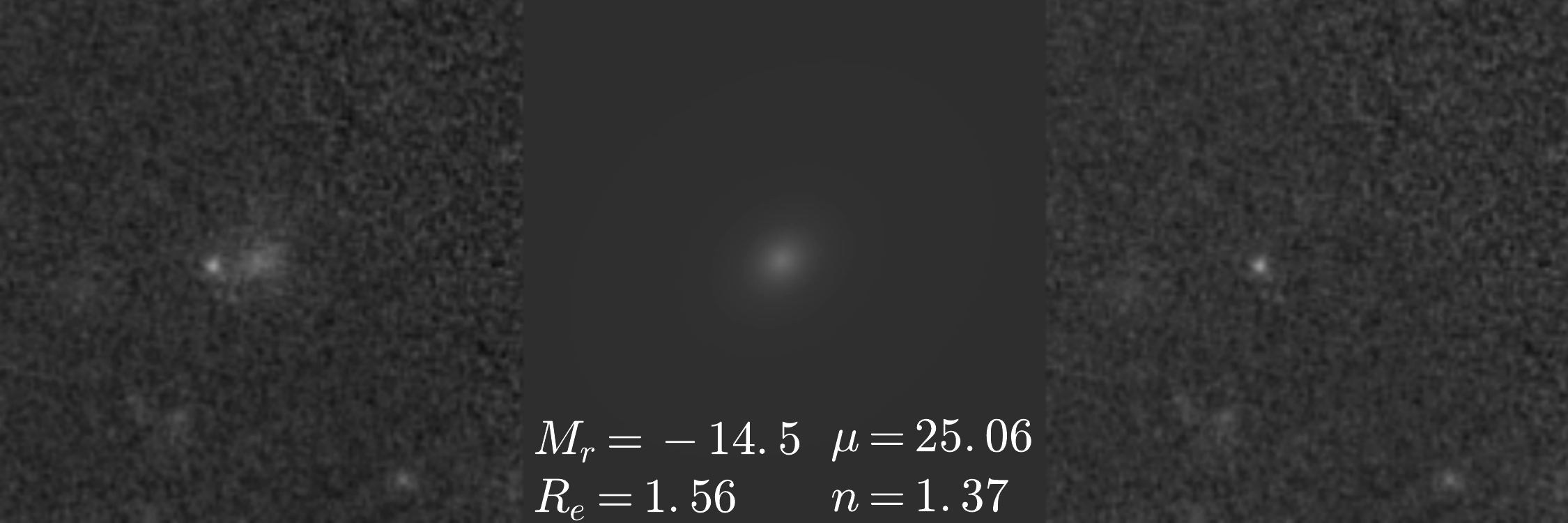}
	\caption{
	Examples of GALFIT fits for six UDGs. For each galaxy, from left to
	right are the F814W image, the GALFIT model and the residual image.
	The best fit S\'ersic parameters are shown, where $M_r$ is the
	absolute $r$-band magnitude, $R_e$ is the circularized effective
	radius in kpc, $\mu$ is the absolute $r$-band mean surface brightness
	within $R_e$ in $\mathrm{mag}~\mathrm{arcsec}^{-2}$, and $n$ is the
	S\'ersic index.  The images are $4.5\arcsec \times 4.5\arcsec$.
	\label{fig:stamps}
	}
\end{figure*}

\begin{enumerate}

\item In the first step of our selection procedure, we conservatively selected
all objects large enough to conceivably be a UDG using the following
\textsc{SExtractor} parameter cuts: $\mathtt{FLAGS} < 4$ (allowing blended
objects and objects with nearby neighbours) and $\mathtt{FLUX\_RADIUS} > 7.4$
pixels, corresponding to $1.0~\mathrm{kpc}$ at $z = 0.308$.

\item UDGs are known to be red (\citealt{vandokkum2015a}; vdB16). So in the
second step we used A2744's red sequence to define a colour cut which allowed
us to isolate the UDG candidates in the A2744 cluster and parallel fields.
This was done by applying a linear fit to the bright end of the
$\mathrm{F814W}-\mathrm{F105W}$ red sequence defined using \textsc{Astrodeep}
\citep{merlin2016,castellano2016} photometric redshifts $0.2 < z_\mathrm{phot}
< 0.4$ and selecting objects with colours between $0.15$ and $-0.5$ of the red
sequence.  No such cut was applied to data from the XDF (and, as will be shown
below, none was needed, as the XDF contains very few UDGs).

\item The next step in our UDG galaxy selection relied on structural parameter
fits to further isolate UDG candidates.  We ran \textsc{GALFIT}
\citep{peng2002} on each candidate to fit a single component S\'ersic model to
each F814W image (F775W for the XDF). We used the \textsc{SExtractor}
segmentation map to mask other detections and models were convolved with a PSF
defined by using \textsc{PSFEx} to produce a model PSF at the location of each
UDG candidate. The resulting effective radii were circularized using $R_{e,c}
= R_e\sqrt{b/a}$. Surface brightness was characterized using
$\langle\mu\rangle_{e,\mathrm{abs}}$, the absolute mean surface-brightness
within $R_e$ \citep{graham2005}. We transformed our surface brightnesses from
F814W (F775W for the XDF) to $r$ assuming a star-formation history given by a
simple stellar population (SSP) with $[\mathrm{Fe}/\mathrm{H}] = -0.6$, an age
of 6.7 Gyr at $z=0.308$, and a \cite{chabrier2003} IMF.  Following vdB16, we
used cuts of $R_{e,c} \geq 1.5~\mathrm{kpc}$, $23.8 \leq
\langle\mu\rangle_{e,\mathrm{abs}} \leq
26.3~\mathrm{mag}~\mathrm{arcsec}^{-2}$ and S\'ersic index $n \leq 4$ to
produce a set of UDG candidates.\footnote{Note that $23.8 \leq
\langle\mu\rangle_{e,\mathrm{abs}} \leq
26.3~\mathrm{mag}~\mathrm{arcsec}^{-2}$ corresponds to $24 \leq
\langle\mu\rangle_e \leq 26.5~\mathrm{mag}~\mathrm{arcsec}^{-2}$ at $z =
0.055$, the mean redshift of the clusters studied in vdB16.} Since UDGs are
round \citep{burkert2016}, we also removed objects with axis ratios $b/a \leq
0.3$ to remove edge-on disks and lensing arcs. A total of 65 UDG candidates
were found in the A2744 cluster field, 63 in the parallel field, and 30 in the
XDF.

\item Each candidate was visually inspected and classified into the following
categories: (i) UDG; (ii) Possible UDG/poorly fit object; (iii) Image
artifact. Most objects in the third category were due to spurious features in
the low signal-to-noise regions at the edges of the frames.

\end{enumerate}

After the final visual inspection, we find a total of 76 UDGs in A2744 (41 in
the cluster field, 35 in the parallel field), while just 4 UDGs are found in
the XDF. All but one of our visually inspected UDGs have a photo-$z$ in the
\textsc{Astrodeep} catalog, and 63 have $z_\textrm{phot} < 1$. The
circularized sizes and mean surface brightness of all objects in our sample
are shown in the left panel of Figure~\ref{fig:galfit}. The black lines show
our size and surface brightness cuts (Step 2 in our procedure above).  For
comparison, we also show the Coma UDGs from \cite{yagi2016} in light grey.
Since the purpose of the XDF observations was to determine the level of
background contamination from field UDGs, the physical sizes of XDF objects
were calculated assuming they are at the same redshift as A2744.  The
right-hand panel of Figure~\ref{fig:galfit} shows the sizes, absolute $r$
magnitudes and stellar masses of A2744 UDG candidates, along with those in
Coma from \cite{yagi2016} for comparison. We calculated stellar masses from
the F814W magnitudes using the same SSP as above.  Examples of six UDGs are
shown in Figure~\ref{fig:stamps}. 

%\vfill
\subsection{Ultra-compact Dwarf Selection}

\begin{figure}
	\includegraphics[width=0.50\textwidth]{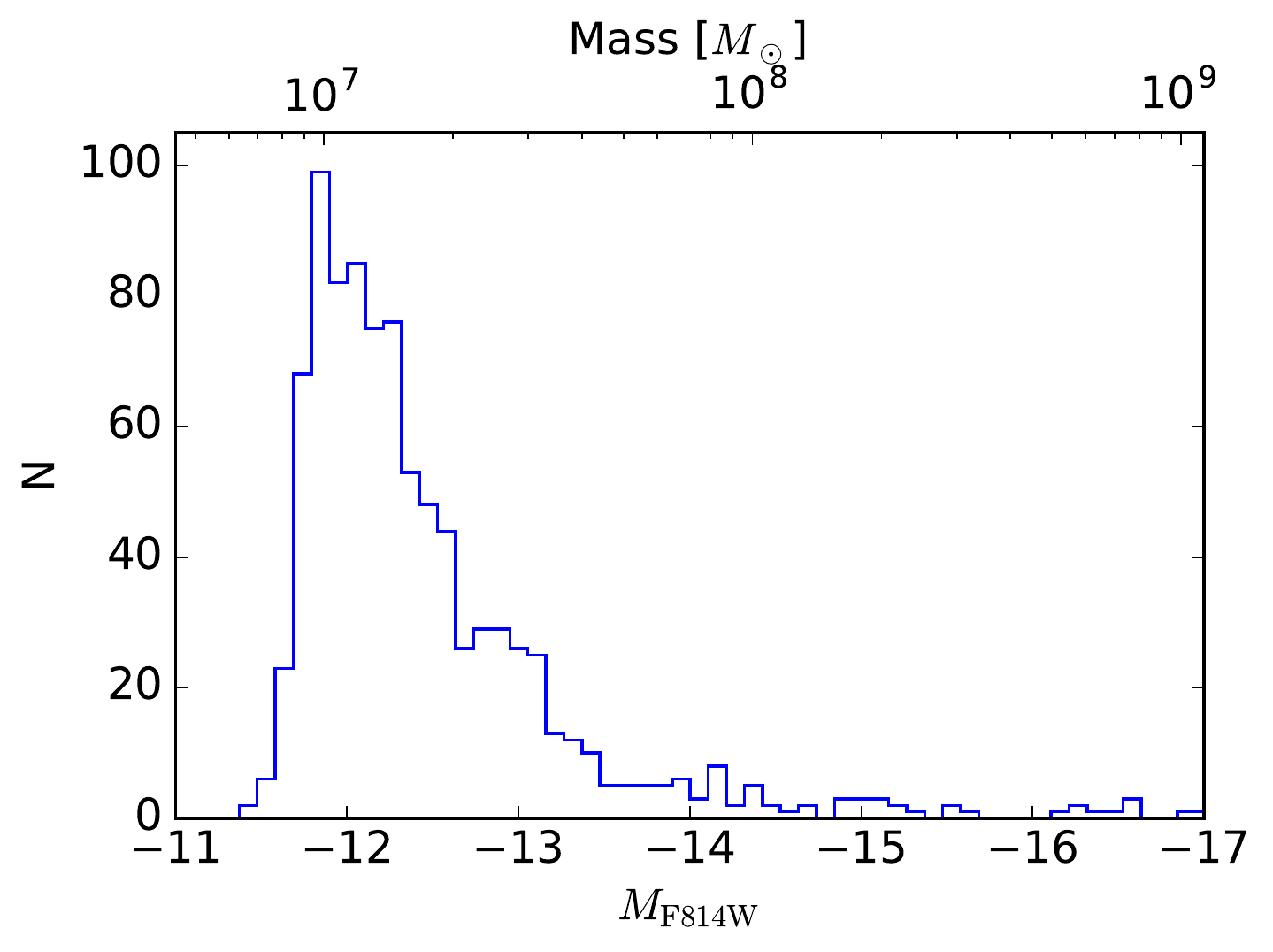}
	\caption{
	Histogram of compact stellar systems in the central 300 kpc of A2744.
	Absolute F814W magnitudes have been converted into stellar masses
	assuming $[\mathrm{Fe}/\mathrm{H}] = -0.6$, old ages and a
	\cite{chabrier2003} IMF. Using a GC upper mass cutoff of $2 \times
	10^6~M_{\odot}$, all of the detected compact systems are UCDs.
	\label{fig:UCD}
	}
\end{figure}

At $z = 0.308$, UCDs are unresolved by \textit{HST}. They are also expected to be
predomininantly found near the brightest cluster galaxies (BCGs).  Therefore,
to detect point sources near the BCGs, we applied a 15 pixel median filter to
the A2744 cluster image and subtracted this off to remove low-frequency power
(e.g.\ from intracluster light and galaxy halos) from the image.
\textsc{SExtractor} was then run in dual image mode using the median filtered
image as the detection image using $\mathtt{DETECT\_MINAREA} = 5$ and
$\mathtt{DETECT\_THRESH} = 1.0$. Point sources were identified on the basis of
image concentration, $C_{3-7}$, given by the difference in an object's
magnitude determined with 3 pixel and 7 pixel diameter apertures.  Point
sources were obtained using the cuts $\mathtt{FLAGS} < 4$ and $C_{3-7} <
1.25~\mathrm{magnitudes}$. Object magnitudes were determined using 4 pixel
($0.12\arcsec$) diameter apertures. An aperture correction of 0.88 magnitudes
was applied by first finding the correction from a $0.12\arcsec$ to a
$1\arcsec$ diameter aperture using our \textsc{PSFEx} PSF, and then correcting
from a $1\arcsec$ diameter to infinity using Table 5 in \cite{sirianni2005}.
The luminosity (mass) distribution of UCD candidates in A2744 is shown in
Figure~\ref{fig:UCD}.

\section{Ultra-diffuse and Ultra-compact Galaxies in Abell 2744}

\begin{figure}
	\includegraphics[width=0.50\textwidth]{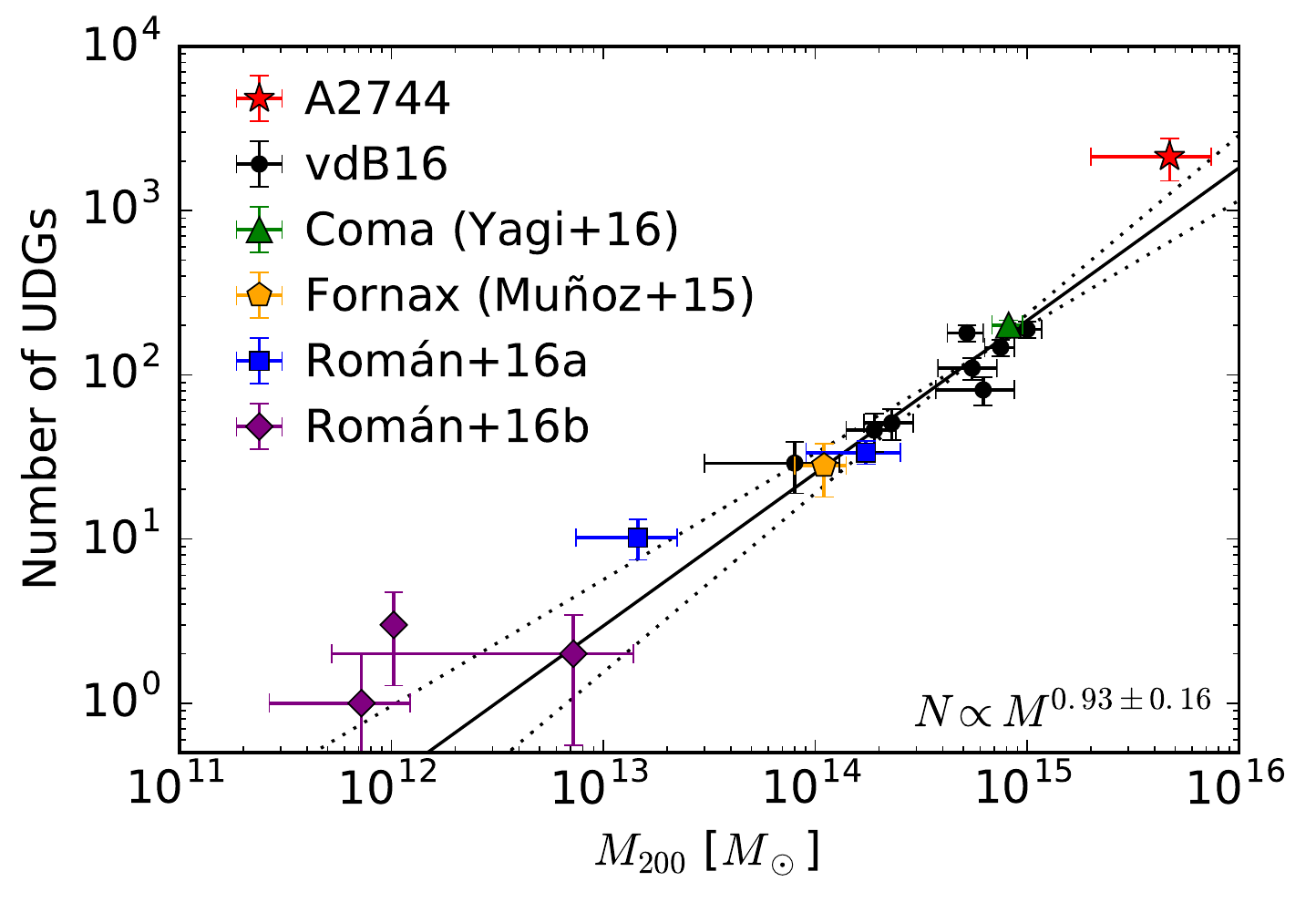}
	\caption{
	Abundance of UDGs with halo mass. We show our estimate of the total
	number of UDGs in A2744 along with values from the literature (see
	text for details). Also shown is the best fit relation from
	vdB16 which has a power-law slope of $0.93 \pm 0.16$. 
	\label{fig:mass_scaling}
	}
\end{figure}

The WFC3 coverage of A2744 and its parallel field contain 76 systems that are
classified as UDGs using the objective criteria noted above. The observations
sample only a small portion of A2744 within $R_{200}$, so this number must be
corrected for geometrical incompleteness. Since, as shown below, the radial
surface density of UDGs appears relatively flat, we simply divide the number
of observed UDGs by the fraction of A2744 observed within $R_{200}$ and
subtract off the expected number of background UDGs in this area. Therefore,
after applying a geometrical and background correction, A2744 contains $2133
\pm 613$ UDGs. This is about 10 times the number that exist in
Coma\footnote{Note that we adopt a considerably more stringent definition for
UDGs than that used by \cite{koda2015}. Using their definition and correcting
for incompleteness yields over 800 UDGs in Coma.}.

Recently, vdB16 showed that the number of UDGs in nearby clusters scales
nearly linearly (in log space) with the mass of the cluster (interior to
$M_{200}$, the number of UDGs scales as $M^{0.93}$).  Adding A2744 ($M_{200} =
5 \times 10^{15}~M_{\odot}$) allows us to extend this relation by 0.7 dex, as
shown in Figure \ref{fig:mass_scaling}, which overplots our A2744 number on
top of the relation of vdB16.  We include UDGs in Coma and Fornax by applying
our selection to the \cite{yagi2016} and \cite{munoz2015} catalogs,
respectively, the numbers in A168 and UGC842 \citep{roman2016a}, and three
Hickson Compact Groups \citep{roman2016b}. For Fornax, the catalog covers the
inner 350 kpc, so we apply a geometrical incompleteness
correction\footnote{\cite{munoz2015} find a flat radial surface density
profile of all dwarfs out to $\sim350$ kpc. We assume UDGs follow the same
profile and that it continues to be flat to $R_{200}$.} out to $R_{200} =
700~\mathrm{kpc}$ \citep{drinkwater2001}.
%Coma ($M_{200} = 8.5 \times 10^{14}~M_{\odot}$)
%Fornax ($M_{200} = 1.1 \times 10^{14}~M_{\odot}$)
A2744 contains about twice the number of UDGs predicted by the 
vdB16 relationship, although the errors are large and the deviation from the
relationship is not significant. 

Recently, \cite{lee2016} studied compact (FWHM $\lesssim$ 400 pc) objects within
the A2744 cluster field (using the parallel field for background subtraction).
These sources are concentrated around the brightest cluster galaxies,
confirming their membership of A2744.  They detected thousands of sources
ranging from a faint limit of around F814W $\sim$ 29.5 to F814W $\sim$ 27. By
fitting a standard globular cluster luminosity function with a peak at F814W =
33.0 (some 3.5 mags below the detection limit) and extrapolating to F814W =
27, they concluded that A2744 contained $147 \pm 26$ UCDs, and a total number
of $385,044 \pm 24,016$ globular clusters.  However, the assumption of a
standard Gaussian GCLF extrapolated to bright magnitudes implies that a
significant number of their GCs have masses greater than $2 \times
10^6~M_{\odot}$ (a widely accepted upper mass cutoff for a GC), and it seems
much more likely to us that the vast majority of the objects identified by
\cite{lee2016} are UCDs.

We note that within 300 kpc of the Fornax cluster centre, the number of UCDs
with masses $> 10^7~M_{\odot}$ is 24 \citep{pfeffer2014}, and similarly in
Virgo, there are 31 \citep{zhang2015}. Scaling by the relative cluster masses
and the predicted relation of Pfeffer et al., one expects between 360 and 720
UCDs in A2744.  This is inconsistent with the 147 UCDs identified by
\cite{lee2016}.  However, our estimate of $385 \pm 32$ (Figure \ref{fig:UCD})
UCDs with masses between $10^7$ and $10^8~M_{\odot}$ within 300 kpc of the
cluster centre\footnote{We use the location of the BCG nearest the X-ray peak
as the cluster centre \citep{owers2011}.} (including a background
correction from the parallel field) lies between these two extremes.

\begin{figure}
	\includegraphics[width=0.50\textwidth]{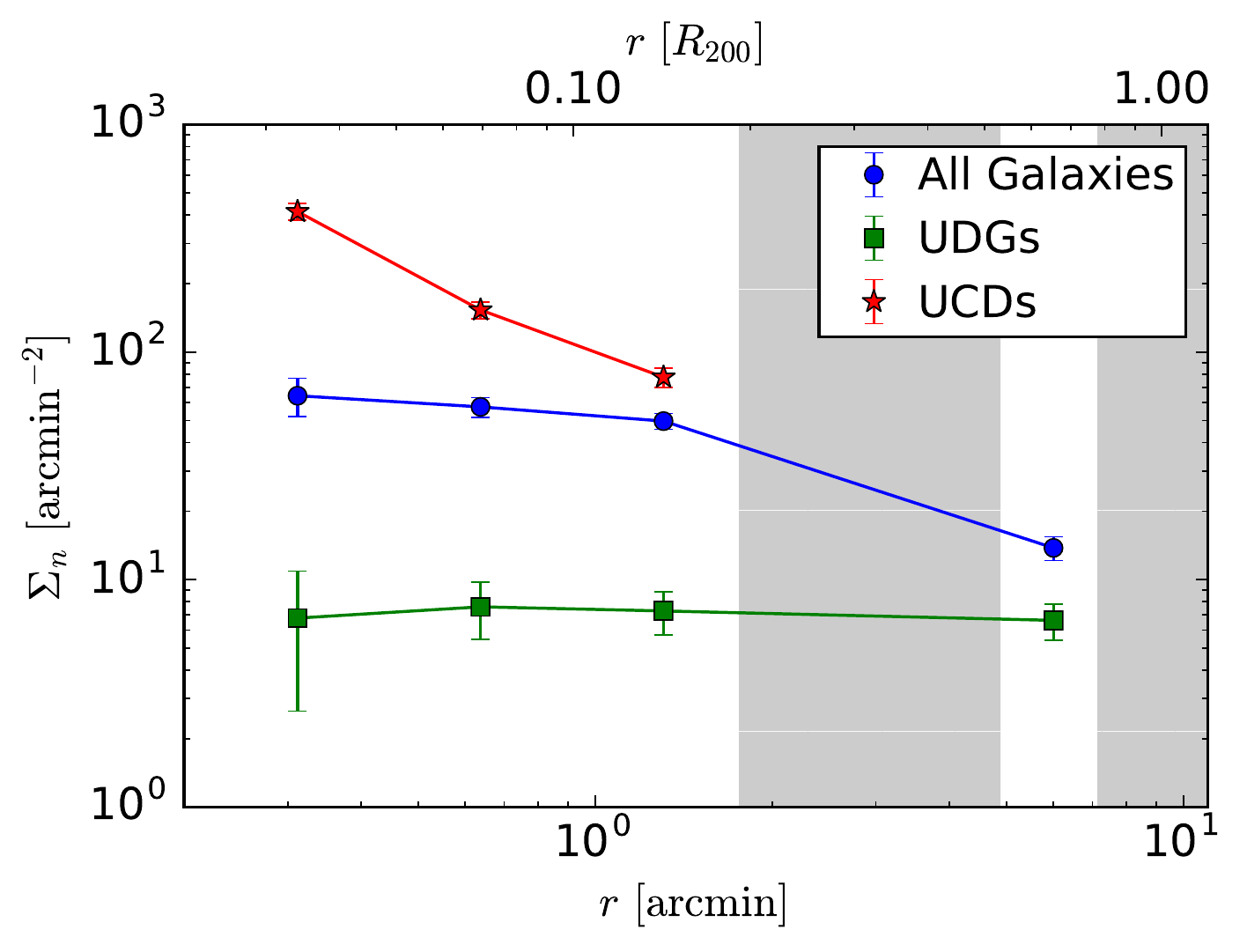}
	\caption{
	Radial surface density distribution of UDGs (green), UCDs (red), and
	\textsc{Astrodeep} \citep{merlin2016,castellano2016} galaxies with
	photometric redshifts $0.2 < z_\mathrm{phot} < 0.4$ and stellar masses
	$> 5\times10^7~M_\odot$ (blue) in A2744.
	A background correction of $0.37~\mathrm{arcmin}^{-2}$ was
	subtracted off the UDG profile (from the XDF), and a correction of
	$76~\mathrm{arcmin}^{-2}$ was applied to the UCD profile (from the
	parallel field).
	The grey regions denote radii not covered by WFC3.
	\label{fig:radial}
	}
\end{figure}

Two UDGs in Virgo, VLSB-A and VLSB-D, appear to be in the process of being
tidally disrupted and host compact nuclei with properties similar to UCDs,
hinting at a transformation from UDG to UCD \citep{mihos2016}. At least one
UDG in A2744 appears to be nucleated (top right of Figure \ref{fig:stamps}).
In addition, the abundance of UCDs is predicted to scale with cluster mass in
a manner similar to that of UDGs \citep[$N_\mathrm{UCD} \propto
M^{0.87}$,][]{pfeffer2014}.  Although the abundance scaling relationships for
UDGs and UCDs appear to be similar, Figure~\ref{fig:radial} shows that UDGs
and UCDs have markedly different radial distributions within the cluster. The
projected surface density distribution of UCDs is very cuspy, rising sharply
toward the centre, whereas the surface density distribution of UDGs is
essentially flat. In fact, vdB16 find the projected surface density of UDGs in
their clusters to be consistent with zero UDGs within a central spherical
region of $r = 0.15 \times R_{200}$.
This points to a picture where some UCDs in A2744 may have once been nuclei or
satellites of infalling UDGs, but that the latter are ultimately destroyed by
tidal forces. As UDGs fall in and dissolve (and, presumably, blend into the
intra-cluster light), they leave behind a residue of unbound, but long lived,
UCDs.

\acknowledgements
Based on observations made with the NASA/ESA Hubble Space Telescope, obtained
from the data archive at the Space Telescope Science Institute. STScI is
operated by the Association of Universities for Research in Astronomy, Inc.\
under NASA contract NAS 5-26555.
These observations are associated with the
Frontier Fields program. We thank NSERC for financial support, and acknowledge
support from the NSF (AST-1616595, AST-1518294, AST-1515084 and AST-1616710).
DF thanks the ARC for financial support via DP130100388 and DP160101608. 

Facilities: \facility{HST (ACS, WFC3)}

\bibliographystyle{apj}

\end{document}